\documentclass{PoS}
\usepackage{amsmath}
\newcommand{\pt}{\ensuremath{\textit{p}_\text{T}\,}}
\newcommand{\raa}{\ensuremath{\text{R}_\text{AA}\,}}
\newcommand{\taa}{\ensuremath{\text{T}_\text{AA}\,}}
\newcommand{\gev}{\ensuremath{\ \text{GeV}\,}}
\newcommand{\gevc}{\ensuremath{\ \text{GeV}/\textit{c}\,}}

\newcommand{\mevc}{\ensuremath{\ \text{MeV}/\text{c}\,}}

\newcommand{\tev}{\ensuremath{\ \rm{TeV}\,}}

\newcommand{\ncoll}{\ensuremath{\langle \text{N}_\text{coll}\rangle \,}}

\newcommand{\sqrs}{\ensuremath{\sqrt{\text{s}}\,}}
\newcommand{\sqrsn}{\ensuremath{\sqrt{\text{s}_\text{NN}}\,}}

\title{The nuclear modification of charged particles in Pb-Pb at $\sqrsn = \textbf{5.02}\,\textbf{TeV}$ measured with ALICE}

\ShortTitle{The nuclear modification factor at $\sqrsn = 5.02 \tev$ }

\author{\speaker{Julius Gronefeld}\thanks{On behalf of the ALICE Collaboration}\\
        GSI Helmholtz Centre for Heavy Ion Research, Darmstadt, Germany\\
        E-mail: \email{j.gronefeld@cern.ch}}

\abstract{
The study of inclusive charged-particle production in heavy-ion collisions provides insights into the density of the medium and the energy-loss mechanisms. The observed suppression of high-\pt yield is generally attributed to energy loss of partons as they propagate through a deconfined state of quarks and gluons - Quark-Gluon Plasma (QGP) - predicted by QCD. Such measurements allow the characterization of the QGP by comparison with models.

In these proceedings, results on high-\pt particle production measured by ALICE in Pb--Pb collisions at $\sqrsn = 5.02\tev$ as well as well in pp at $\sqrs=5.02\tev$ are presented for the first time.
The nuclear modification factors (\raa) in Pb--Pb collisions are presented and compared with model calculations.
}

\FullConference{Fourth Annual Large Hadron Collider Physics\\
		13-18 June 2016\\
		Lund, Sweden}

\begin{document}

\section{Introduction}

Transverse momentum spectra measurements at RHIC at $\sqrsn = 200 \gev$ \cite{rhicphenix,rhicstar} have shown that charged-particle yields in heavy-ion collisions are suppressed compared to a superposition of independent nucleon-nucleon collisions (binary collision scaling). 
This observation is related to parton energy loss in the hot and dense QCD matter created in the collision of heavy ions, leading to a modification of transverse-momentum (\pt) distributions of the resulting particles, as initially suggested by Bjorken in 1982 \cite{bjorken}.\\
Recent results from ALICE \cite{raa276} show that hadron yields at high \pt in central Pb--Pb collisions at $\sqrt{\text{s}_\text{NN}} = 2.76\,\tev$ are suppressed even stronger than at RHIC, indicating a hotter and denser medium. This suppression is present up to very high \pt and can also be seen in jets \cite{jets}.\\
To rule out nuclear matter effects as the origin ot the observed suppression in heavy-ion collisions control measurements with p--Pb collisions were performed. In this system no suppression was observed and binary collision scaling was found to be valid for intermediate to high-\pt particle production \cite{ppb}.\\
The suppression is often quantified in terms of the nuclear modification factor:

\begin{displaymath}
	\raa(\pt) =\frac{1}{\langle \taa \rangle} \frac{dN_\text{AA}/d\pt}{d\sigma_{pp}/ d\pt}.
\end{displaymath}

\noindent Here, $dN_\text{AA}/d\pt$ represents the \pt-differential charged-particle yield in nucleus-nucleus (AA) collisions, while $d\sigma_{pp}/ d\pt$ stands for the \pt-differential cross section in proton-proton (pp) collisions. The average nuclear overlap function $\langle \taa \rangle$ is determined by Glauber Monte-Carlo calculations for each class of centrality. It relates to the number of binary collisions \ncoll ($\langle \taa \rangle =  \ncoll / \sigma_{inel}^{NN}$) and is strongly dependent on the collision centrality.
In absence of medium effects the nuclear modification factor will be equal to unity, while $\raa < 1$ indicates a suppression of charged-particle yields compared to binary collision scaling. \\
\indent In previous measurements by ALICE at $\sqrsn =2.76 \tev$\cite{raa276}  the suppression is observed to be stronger for central collisions and less pronounced in peripheral collisions. \raa also shows a characteristic dependence on \pt, reaching a minimum of $\sim 0.1$ at $\pt \approx 7 \gevc$ for most central collisions and a rise to $\sim 0.4$ at high \pt, indicating a strong suppression even for $\pt \approx 50\gevc$. 
Increasing the center of mass energy from $2.76 \tev$ to $5.02 \tev$ per nucleon pair the number of charged particles increases by of 20\% \cite{nch} suggesting the formation of a medium with a larger volume, longer lifetime or higher temperature.

\section{Analysis}

Primary charged particles are defined as all prompt charged particles produced in the collision including all decay products, except for products from weak decays of light flavor hadrons such as $K^{0}_{S}$ and $\Lambda$.
Events are triggered using two forward scintillator detectors, V0A and V0C, which cover the full azimuth and pseudorapidity range of $2.8<\eta<5.1$ and $-3.7<\eta<-1.7$, respectively. Charged particles are reconstructed using the ALICE Inner Tracking System (ITS) as well as the Time Projection Chamber (TPC). Both subdetectors cover the full azimuth and pseudorapidity of $|\eta|<0.9$. In Pb--Pb collisions the sum of amplitudes measured in V0A and V0C is used to estimate the collision centrality \cite{centrality}. A detailed description of the experimental setup and the reconstruction procedure can be found in \cite{aliceperformance}.\\

\begin{figure}[h]
	\centering
	\includegraphics[width=0.47\textwidth, height=9.8cm, clip=true, trim=1cm 0 0 1cm]{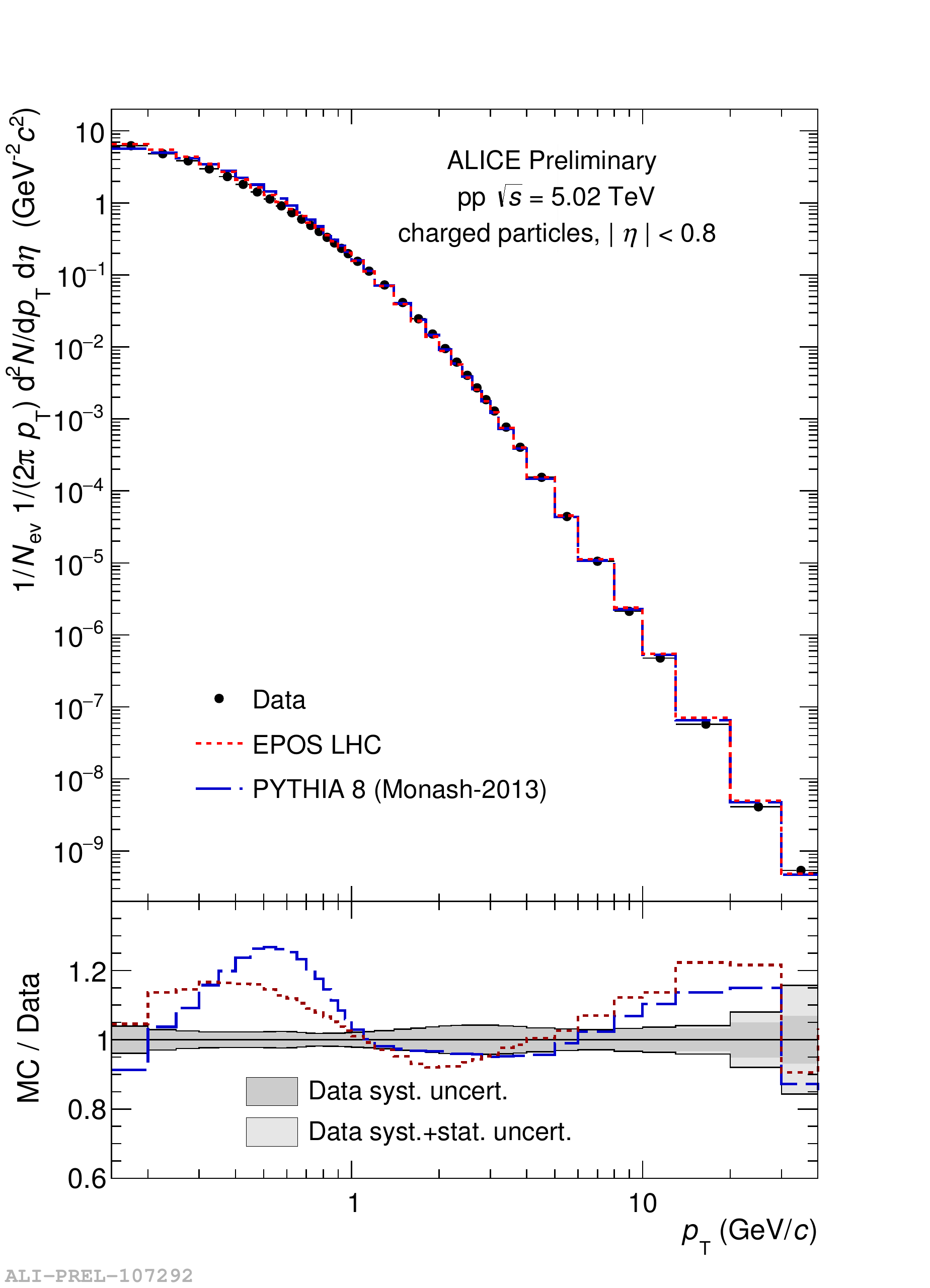}
	\includegraphics[width=0.51\textwidth, height=9.8cm, clip=true, trim=0 0 1cm 1cm]{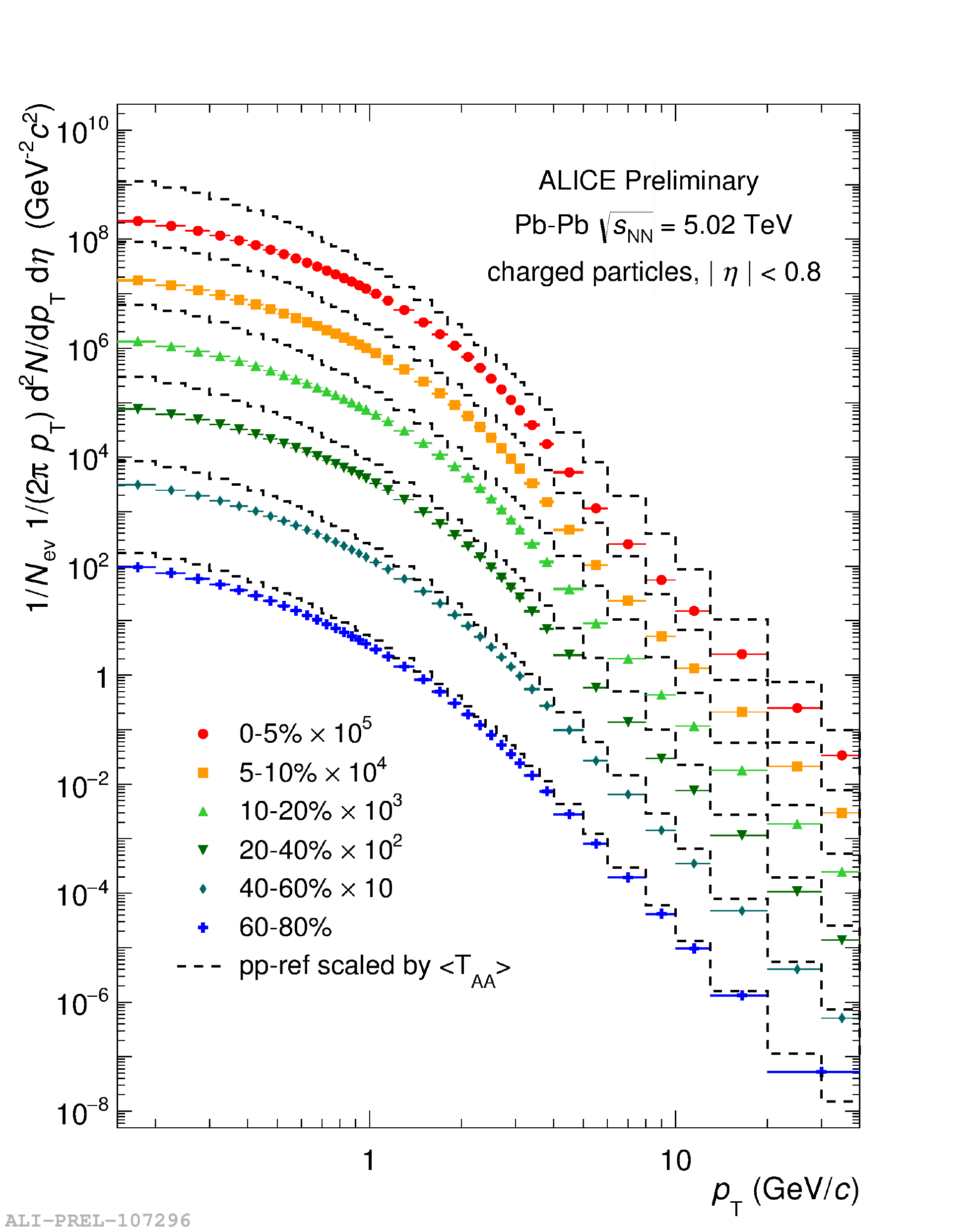}
	\caption{
	\textbf{Left:} Transverse momentum distribution of charged particles measured in  pp collisions compared to model calculations \cite{pythia,epos}. The lower panel shows the ratio of model calculations to the measured spectra. The measurements uncertainties are shown as a band around unity.
	\textbf{Right:} Transverse- momentum distribution of charged particles measured in Pb--Pb collisions for different classes of centrality. The dotted curves represent the measured \pt distribution in pp collisions scaled by the nuclear overlap functions (\taa).
	}
	\label{combspec}
\end{figure}

The analysis presented is based on a data sample recorded in late 2015 containing $3.3 \cdot 10^{6}$ Pb--Pb collisions and $25 \cdot 10^{6}$ pp collisions, representing 3\% and 25\% of the total collected statistics, respectively. The data are corrected for geometric acceptance and reconstruction efficiency using simulations based on GEANT3 \cite{geant} in combination with HIJING \cite{hijing} for Pb--Pb and PYTHIA8 \cite{pythia} for pp collisions.
The current track selection was improved by rejecting tracks not reaching a minimal track length in the active volume of the TPC.

To obtain the charged-particle yield as a function of $\textit{p}_\text{T}$, corrections are made for tracking efficiency and acceptance ($\sim 70\%$), for contamination by secondary particles from weak decays or secondary interactions ($\sim 10\%$, important at low \pt) and for \pt-resolution ($\sim 2\%$, important at $\pt > 20\gevc$).
To account for differences in the particle composition of event generators and the data, the charged-particle reconstruction efficiency was calculated from the particle-dependent efficiencies weighted by the relative abundances of each particle measured in pp at 7\tev and Pb--Pb at 2.76\tev.
The correction for contamination with secondary particles is taken from Monte-Carlo simulations. 
To compare data and simulation the differences in the closest approach of tracks to the event vertex (DCA) between data and simulation are investigated. 
It is found that the contamination correction from Monte Carlo has to be scaled up by $\sim 30\%$ to match the data.

The total relative systematic uncertainties are in the range range 3.3-7\% for pp and 2-6\% for central Pb--Pb collisions (0-5\%) and 4-5.5\% in peripheral Pb--Pb collisions (60-80\%). The uncertainties are dominated by the uncertainties due to the track selection as well as the uncertainty on the tracking efficiency and secondary-particle rejection. For peripheral Pb--Pb collisions the centrality estimation also contributes with $\sim 3\%$ to the total uncertainties. Overall, the systematic uncertainties in the current analysis were reduced by about $50\%$ compared to previous analyses, owing to an improved reconstruction and calibration procedure in Run2, as well as to improved track selection methods. 

\begin{figure}
	\centering
	\vspace{-0.5cm}
	\includegraphics[width=0.85\textwidth]{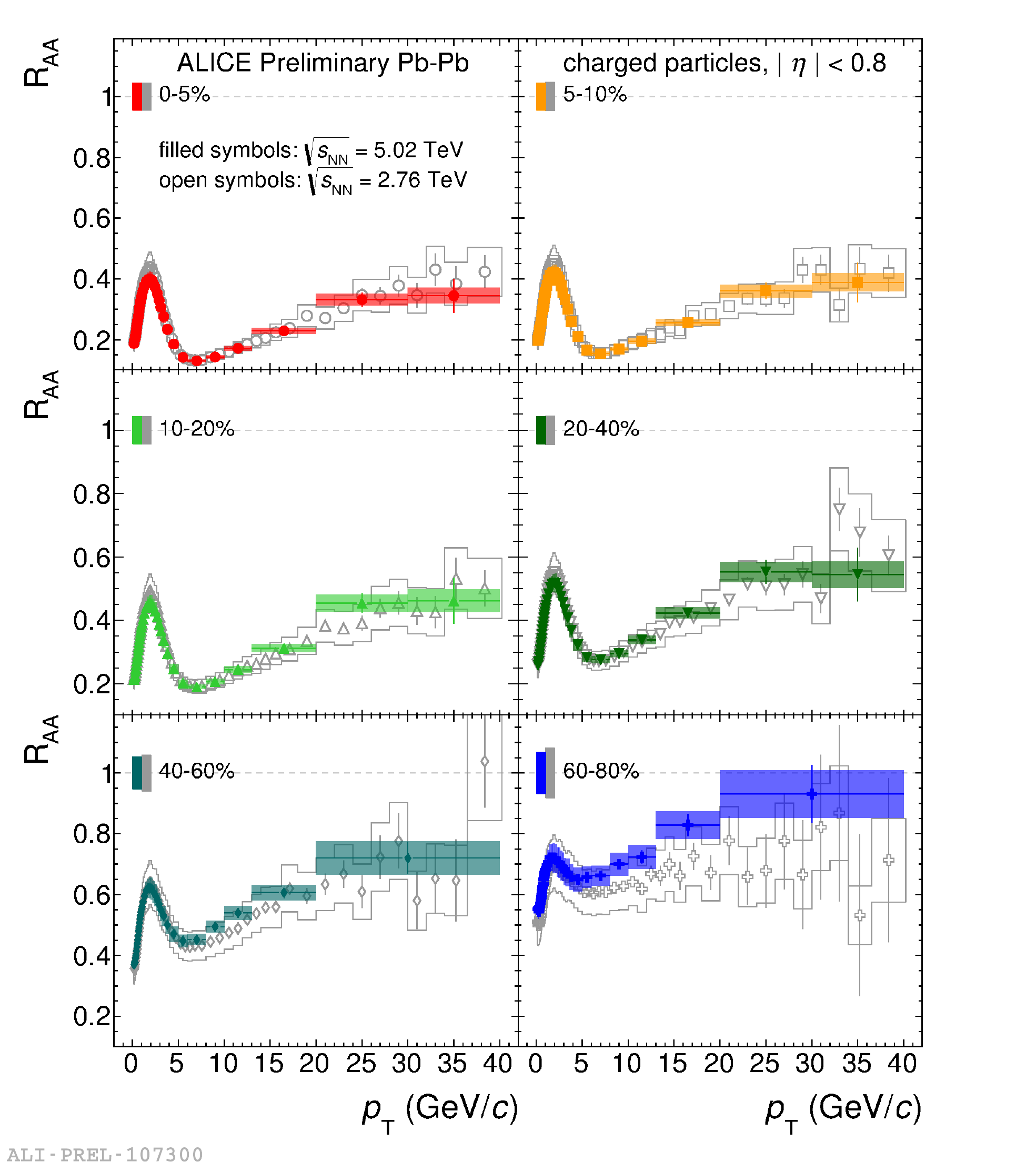}
	\caption{The nuclear modification factor \raa as a function of transverse momentum, $\textit{p}_\text{T}$, for different centrality classes in Pb--Pb collisions. The filled symbols represent the measurement at $\sqrt{\text{s}_\text{NN}} = 5.02\tev$ while the open symbols stand for the measurement at $\sqrt{\text{s}_\text{NN}} = 2.76\tev$ \cite{raa276}. The normalisation uncertainty is shown for both energies as a box around unity.}
	\label{raa}
\end{figure}

\clearpage
\section{Results}
\begin{figure}
	\centering
	\vspace{-0.9cm}
	\includegraphics[width=0.49\textwidth]{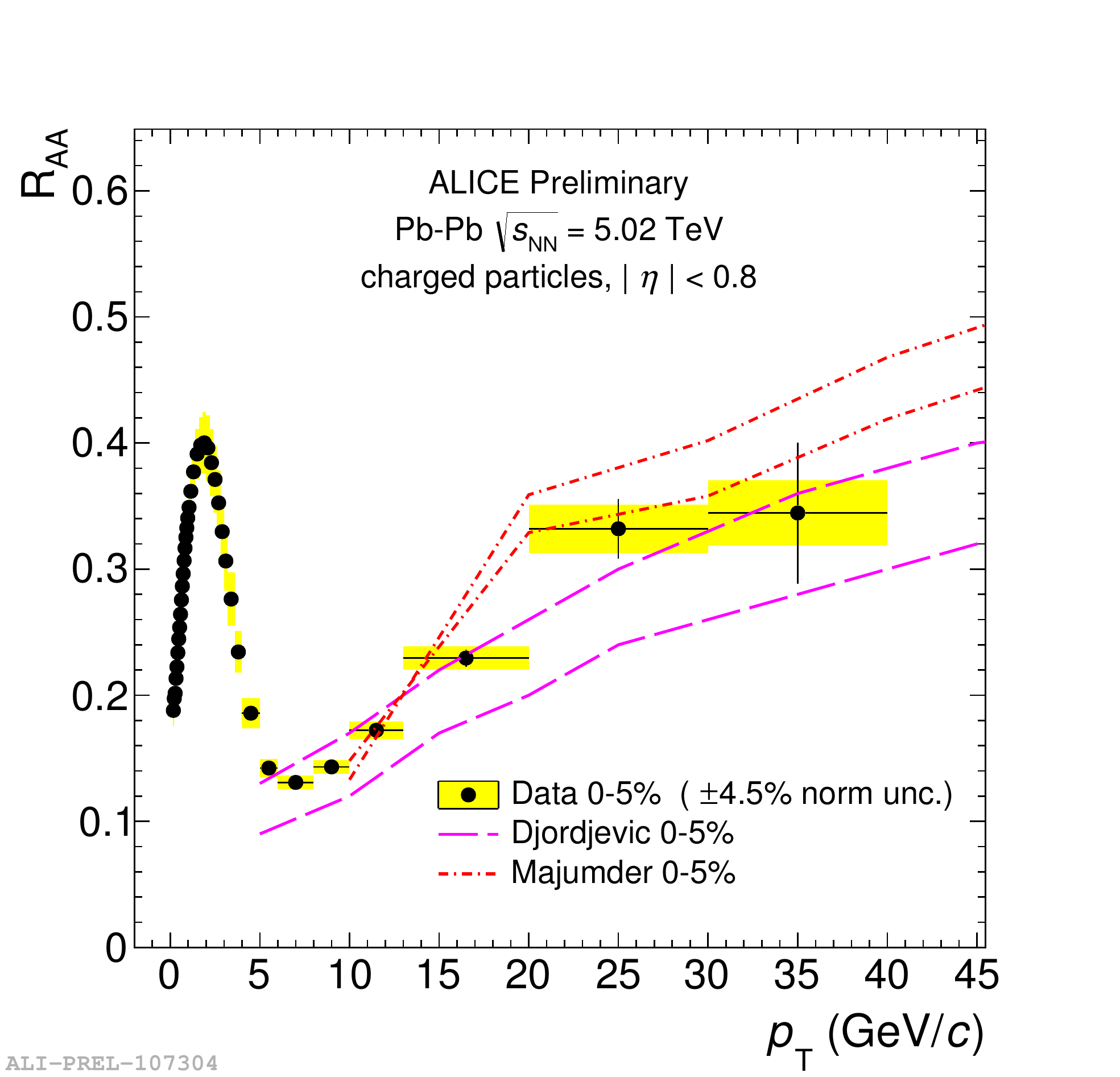}
	\includegraphics[width=0.49\textwidth]{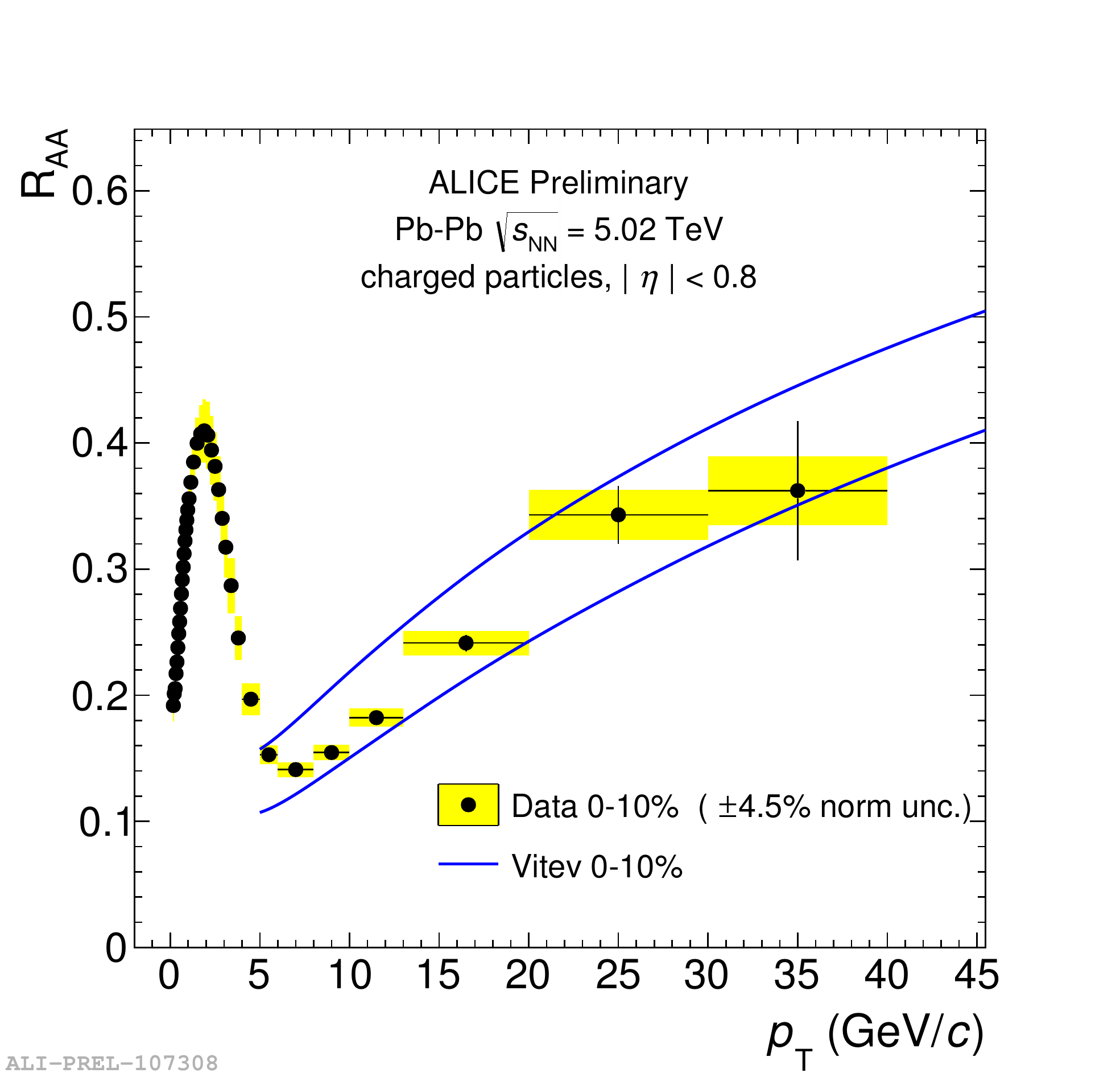}
	\caption{The \raa measured at $\sqrsn = 5.02\tev$ for 0-5\% (left) and 0-10\% centrality (right) compared with model calculations \cite{djordjevic,majumder,vitev}.}
	\label{theory}
\end{figure}

Figure\ \ref{combspec} (left) shows the transverse momentum distribution of charged particles in pp collisions at $\sqrt{\text{s}}=5.02 \tev$. The measurement was performed in the kinematic range of $150 \mevc \leq \pt < 40 \gevc$ for particles with a pseudorapidity of $|\eta|<0.8$. The measurement is compared to predictions from  the EPOS event generator \cite{epos} and PYTHIA8 with the Monash-2013 tune \cite{pythia}. Both event generators describe the measurement at intermediate \pt, however overestimating the cross section for low and high transverse momenta.\\
\indent The transverse momentum distribution of charged particles from Pb--Pb collisions is shown in Fig.\  \ref{combspec} (right) for six classes of centrality. The spectra are compared with the pp measurement scaled by the nuclear overlap function. Comparing the spectra one observes that the \pt distribution in peripheral collisions is similar to the scaled spectrum in pp collisions while the difference increase towards more central collisions.\\
\indent The nuclear modification factors as a function of \pt for six centrality intervals are shown in Fig.\  \ref{raa}. The filled symbols represent the measurement at $\sqrsn = 5.02\,\text{TeV}$, while the open symbols represent the published measurement at $\sqrsn = 2.76\tev$ \cite{raa276}. Both measurements exhibit similar features, showing only moderate suppression ($\raa \sim 0.6-0.7$) for peripheral collisions (60-80\%). In more central collisions, a pronounced minimum at about $\pt \sim 6-7 \gevc$ develops, while for \pt > 7 GeV/c there is a significant rise of the nuclear modification factor. Within their systematic and statistical uncertainties the measurements at 2.76 \tev and 5.02\tev agree. Having in mind that the spectra tend to harden for higher center-of-mass energies, this agreement could hint towards a stronger energy loss in a hotter and denser medium.\\
\indent The nuclear modification factor for central collisions is compared to model predictions in Fig.\  \ref{theory}. All model predictions succeed in describing the high-\pt rise of the nuclear modification factor within its uncertainty. The predictions of Djordjevic et al. \cite{djordjevic} and by Majumder et al. \cite{majumder} are in agreement with the measured \raa for 0-5\% centrality (Fig.\  \ref{theory} left panel), the \raa for 0-10\% centrality (right panel) is well described by calculations of Vitev et al. \cite{vitev}.

\clearpage
\section{Summary}
The strong suppression of charged-particle yields observed in Pb--Pb collisions at $\sqrt{\text{s}_\text{NN}} = 5.02\tev$ exhibits the same characteristics as at $\sqrt{\text{s}_\text{NN}} = 2.76\tev$. A strong suppression is observed in central collisions, with the maximum suppression around $6-7\gevc$ becoming smaller towards higher \pt. Compared to the previous measurement it was possible to significantly reduce the systematic uncertainties. A comparison of data to predictions shows that all investigated models succeed in describing the \raa within its uncertainty.\\
\indent ALICE is currently analyzing the full 5.02\tev dataset.
In addition, the improved analysis method will be extended to the $\sqrsn = 2.76 \tev$ dataset, in order to further improve the precision in the comparison of the results at the two collision energies.

\end{document}